\documentclass[natbib,namedreferences]{solarphysics}
\usepackage[optionalrh]{spr-sola-addons} 
\usepackage{graphicx}        
\usepackage{color}           
\usepackage{url}             
\usepackage{aas_macros}

\begin{document}

\begin{article}

\begin{opening}

\title{Emission of Type II Radio Bursts - Single-Beam versus Two-Beam Scenario}

%
\author{U.~Ganse$^{1}$\sep
				P.~Kilian$^{1}$\sep
        R.~Vainio$^{2}$\sep
        F.~Spanier$^{1}$
       }

%

%
  \institute{$^{1}$ Lehrstuhl f\"ur Astronomie, Universit\"at W\"urzburg, Germany
                     email: \url{uganse@astro.uni-wuerzburg.de} \\
             $^{2}$ Department of Physics, University of Helsinki \\
             }

\begin{abstract}
	The foreshock region of a CME shock front, where shock accelerated electrons
	form a beam population in the otherwise quiescent plasma is generally assumed
	to be the source region of type II radio bursts. Nonlinear wave interaction
	of electrostatic waves excited by the beamed electrons are the prime
	candidates for the radio waves' emission.

	To address the question whether a single, or two counterpropagating beam
	populations are a requirement for this process, we have conducted
	2.5D particle in cell simulations using the fully relativistic ACRONYM code.

	Results show indications of three wave interaction leading to electromagnetic
	emission at the fundamental and harmonic frequency for the two-beam case. For
	the single-beam case, no such signatures were detectable.
\end{abstract}

%

\end{opening}

%
\section{Introduction}\label{s:intro} 

Solar radio bursts have been observed since the earliest days of astronomical
radio observations. Based on spectral morphology and development timescales,
they are classified into 5 types \citep{WildMcCready}. While suitable models
for the emission of most types of radio bursts have since been found, an
explanation of type II, and to some extent type III radio bursts' emission
mechanisms remains elusive.

\begin{figure}[htb]
	\centerline{\includegraphics[width=0.8\textwidth,clip=]{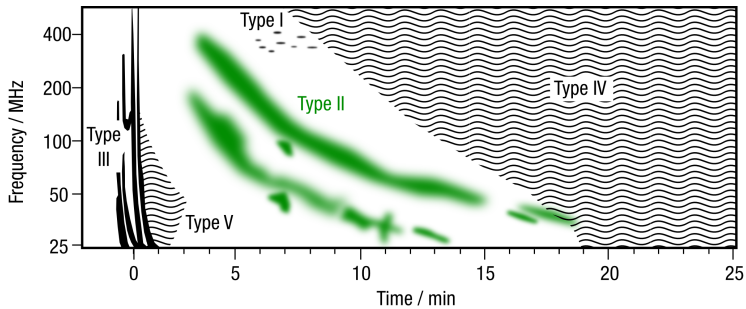}}
	\caption{Exemplary classification of solar radio bursts by their morphology in a dynamic spectrum.
	Type II bursts are characterized by their two-banded emission of fundamental
	and harmonic frequency, which is slowly drifting towards lower frequencies.}\label{fig:dynspectrum}
\end{figure}

The morphology of type II bursts typically shows a two-band emission
spectrum, consisting of the \emph{fundamental} emission band and of the
\emph{harmonic} emission band at about twice the frequency of the fundamental
\citep[in a few, near-limb events signals of third harmonic emission are also
discernable, see][]{ThirdHarmonic}. The fundamental frequency is believed
to correspond to the plasma frequency of the emission region, slowly decreasing
over time as the coronal/interplanetary shock travels outwards into the heliosphere.
\citep{Cane1987, NelsonMelrose}

Statistical studies have provided firm evidence of a correlation between type
II bursts and the propagation of shock fronts outwards from the
sun \newline\citep{NelsonMelrose, WorkingGroupD}. These are in many cases driven by
coronal mass ejections (CME), but have also been observed as blast waves from
flare energy release \citep{TypeIIFlare}. It is therefore reasonable to assume
that electron processes at the shock are responsible for the emission \citep{Mann1995}.
At the time of radio emission, a CME shock front can span from the low solar
chromosphere outwards to multiple solar radii, thus encompassing a large number of
density scales and hence, potential plasma frequencies \citep{Pomoell}. The
observed radio emissions, on the other hand, are quite narrowband.
Consequently, the emission region can not be spread out along a large area of
the shock front, but is
rather required to have a small spatial scale, confined to a small subset of
the shock \citep{schmidtCMEshocks}.  The emission region is thus expected to
have special properties setting it apart from the rest of the shock.

\subsection{Type III Bursts}\label{s:theory}

Very similarly to type II bursts, the so-called type III bursts also feature
multi-banded emission, albeit with a much faster drift towards low frequencies
and hence shorter total burst duration.

For these bursts, conclusive observational evidence exists for a connection to
impulsive electron release in solar flares. 
As the accelerated electrons travel outwards on open field lines, they form an
electron beam population within the solar wind plasma, which can excite Langmuir waves
through Cherenkov-type instabilities \citep{KarlickyVandas, Tsiklauri2010}.

These Langmuir waves ($L$) are then assumed to take part in three wave
interaction processes \citep{Melrose}, leading to emission of transverse
electromagnetic waves ($T$) and sound waves ($S$):
\begin{eqnarray}
	L &\rightarrow& L' + S \label{eq1} \\ L
	&\rightarrow& S + T(\omega_{pe}) \label{lts}\\ 
	L + L' &\rightarrow& T(2\omega_{pe})
	\label{llt} \label{melrosecouplings}
\end{eqnarray}

\begin{figure} 
	\centerline{\includegraphics[width=0.47\textwidth,clip=]{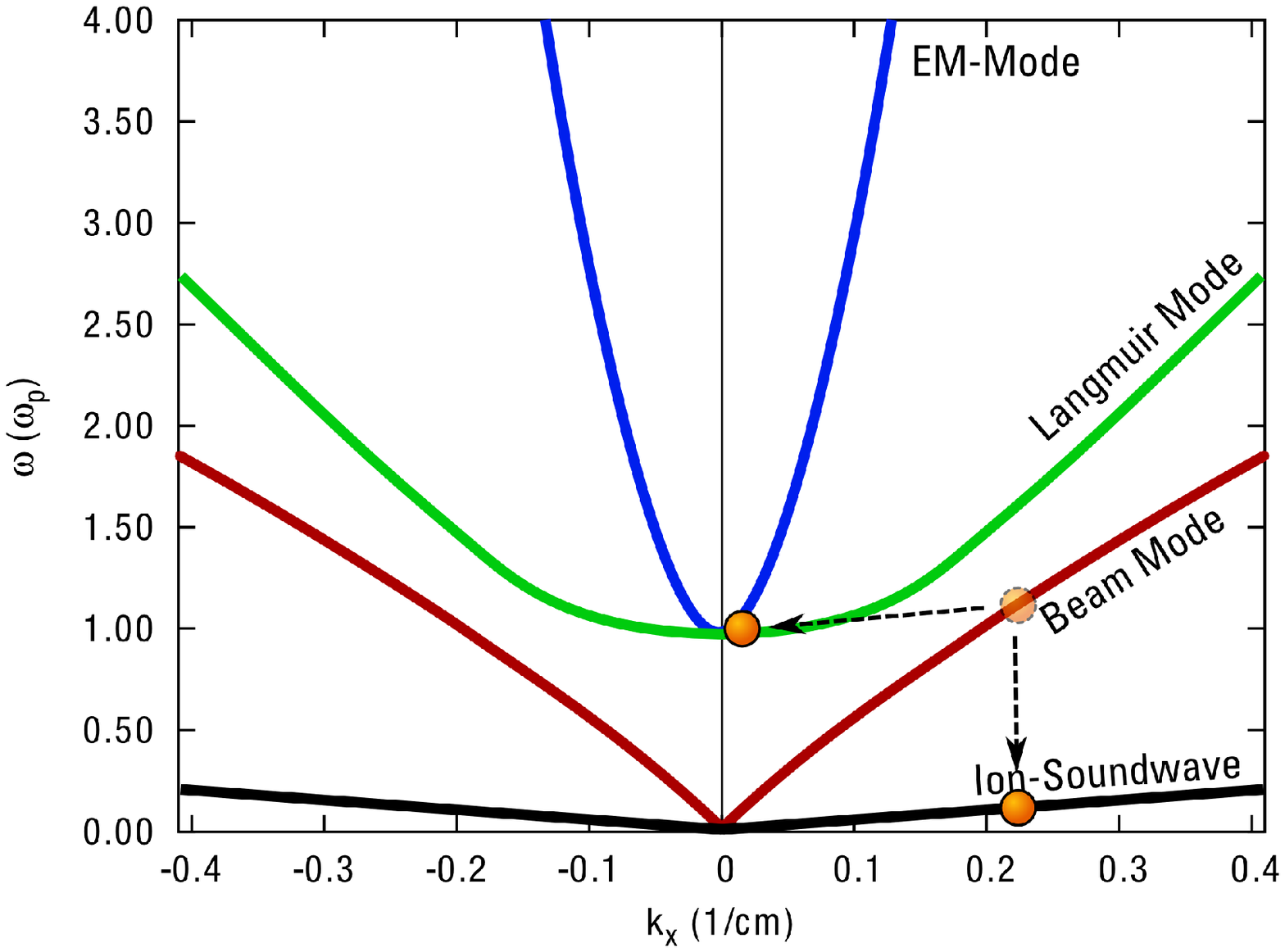}\hspace{1em}\includegraphics[width=0.47\textwidth,clip=]{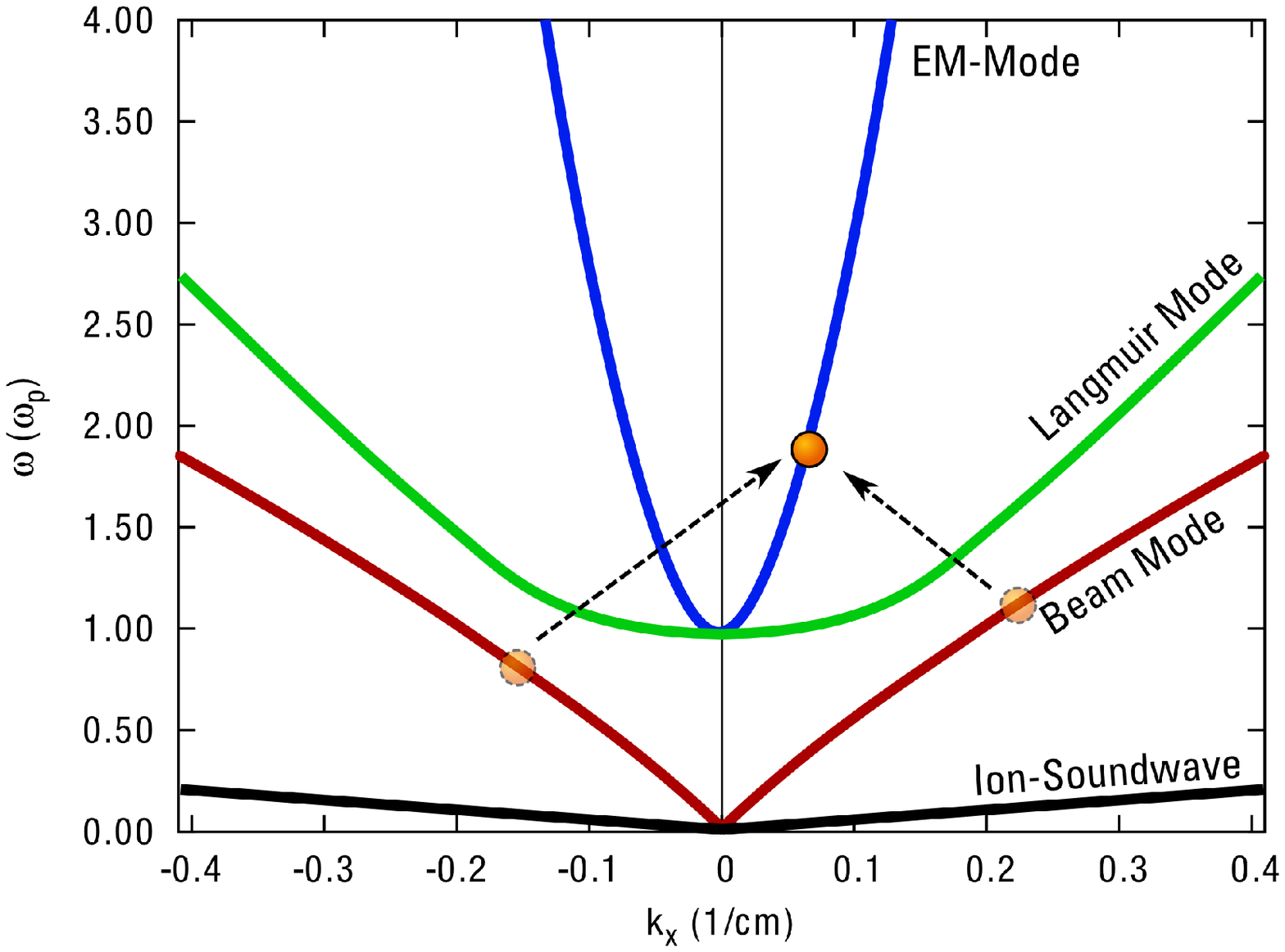}}
	\caption{Sketch of three wave interaction processes believed to be the origin of
	fundamental (left) and harmonic (right) emission of type II radio bursts. Decay
	of beam-driven longitudinal waves into sound-waves and electromagnetic waves at
	the plasma frequency akin to the normal plasma emission process is responsible
	for the fundamental emission, whereas a coalesce process of two
	counter-directed modes leads to harmonic emission.}\label{fig:dingsprozesse}
\end{figure}

In a more detailed analysis, \cite{WillenGeneralizedLangmuir} treated plasma
wave behaviour in the presence of a beam population, and identified not the
Langmuir wave, but the so-called \emph{beam driven mode} as a possible source
of three-wave interactions.
As the beams' intensity decreases, energy in this mode would be transferred
into the background plasma's eigenmodes, thus leading to similar processes as
outlined in eq. \ref{eq1}-\ref{llt}, illustrated in figure
\ref{fig:dingsprozesse}.

Experimental proof of these processes has been challenging: satellite data
\citep{Pulupa2007}, being intrinsically point-wise, can not yield information
about the momenta and energies of participating waves in three-waves
interactions.

While numerical investigation of three-wave interaction has been successfully
performed using MHD codes \citep{Wisniewski}, that technique is not applicable
to the kinetic waves participating in the interactions outlined here.
Particle-in-cell codes on the other hand, which are the tool of choice for
kinetic simulations, have been too numerically expensive to be able to treat
this problem self-consistently in the past. With the exponential growth of
computing power, the length- and timescales of these emission models are coming
into the range of top-end supercomputers now.

\section{Emission Model}\label{s:model}

Due to the similarity to type III bursts in spectral shape, it seems reasonable
to assume that electron beams are likewise responsible for type II radio
emission. Assuming that the electron acceleration on the shock front is
localized to a specific magnetic field configuration also elegantly solves the
problem of the emissions' narrow-bandedness.

\begin{figure}[htb]
	\centerline{\includegraphics[width=0.9\textwidth,clip=]{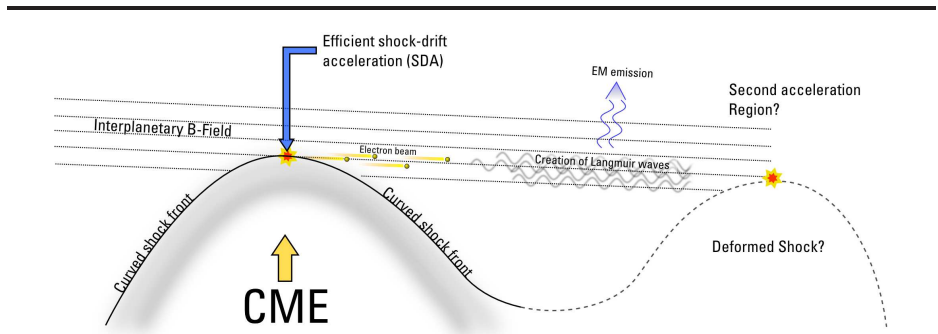}}
	\caption{Sketch of the large scale model of the proposed emission process of
	type II radio bursts.  Electrons accelerated through shock drift acceleration
	form a beam population in the foreshock plasma, exciting Langmuir waves which
	then undergo three-wave interactions.  A variant of this model allows for
	multiple sites of electron acceleration on a deformed shock (such as a shock
	ripple) injecting counterstreaming electron beams into the same region.
	}\label{fig:largescale}\label{fig:counterstream}
\end{figure}

In the predominant model, \cite{Holman1983} proposed that electron shock drift
acceleration on a curved shock front is the source of these electron beams. 
Following the interplanetary field from the point where the shock is exactly
perpendicular into the foreshock plasma, these electron
beams then excite Langmuir waves and undergo three-wave interactions much like
eq.\ \ref{eq1}-\ref{llt} predicts for type III bursts. Since the efficiency of
drift acceleration increases sharply around the sites where the shock angle is
close to $90^\circ$, the source region of the electron beams is confined to a
small area on the curved shock, thus explaining the narrowband nature of the
emission.

An open question in these models remains whether the process of equation
\ref{eq1} itself -- that is the decay of a forward-directed Langmuir wave into
a forward-directed sound wave and a backward-directed Langmuir wave -- is
sufficient to produce a large seed population of counterpropagating Langmuir
waves for the process of equation \ref{llt} to produce significant intensities
of radio emission at twice the plasma frequency.
Alternatively, counterstreaming electron beams, directly exciting
counterpropagating Langmuir waves may be required, as depicted in figure
\ref{fig:counterstream}. These might be created in the case of shock ripples,
or deformations of a shock due to shock-shock interaction, as reported multiple
times in cases of very strong type II bursts. 

\section{Numerical Simulation}\label{s:numerics}

In order to decide whether a two beam scenario is required for harmonic
emission, or whether a single-beam scenario is sufficient, we have conducted
simulations using the ACRONYM particle in cell code.

\subsection{ACRONYM Code}
The ACRONYM code, developed and maintained at the department of Astronomy, University of W\"urzburg,
is a fully relativistic, 2nd order particle-in-cell code for astronomical,
heliospheric and laboratory plasmas. Using MPI-parallelization, the code is
running on all major supercomputer platforms \newline\citep{acronym11}. 

For the simulations in this project, computing time on Jugene (BlueGene/P) at
the J\"ulich Supercomputing Centre and Louhi (Cray XT-5) at the CSC Helsinki
was employed. The runs took a total of about 1 Mio.\ CPU-hours on these machines.

\subsection{Setup}
Since a particle in cell code carries the intrinsic constraint of resolving the
plasma debye length (in the case of solar wind, $\lambda_D \approx
1\,\mathrm{cm}$), the maximum size of a simulation domain is severely limited
to the resolution of plasma-microphysical phenomena. A simulation of the
complete shock environment depicted in figure \ref{fig:largescale} would be
prohibitively expensive in terms of computing power. The simulations rather
focus on the beam-plasma interaction and resulting three-wave processes in the
foreshock plasma. 

\begin{figure} 
	\centerline{\rotatebox{270}{\includegraphics[height=0.8\textwidth,clip=]{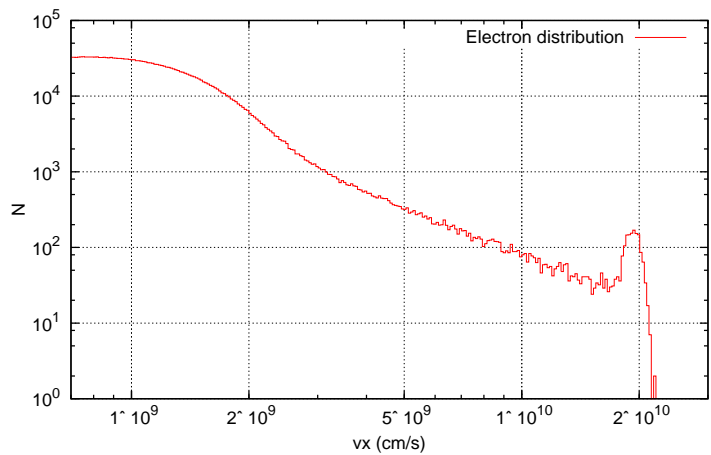}}}
\caption{Electron velocity ($v_x$) distribution at the start of a simulation in
the single-beam case. A second, symmetric beam is included in the two-beam
case}\label{fig:setup}
\end{figure}

The simulations are set up as 2.5D (2D3V) rectangular grids with periodic boundaries,
which are homogeneously filled with the background foreshock plasma under
quiescent solar wind conditions ($T\approx 0.5 \,\mathrm{MK}$, $\rho=2.5\cdot10^{7}
\mathrm{cm}^{-3}$, $B=1 \mathrm{G}$), with a thermal particle distribution. On
top of that, either a single or two counterstreaming beamed electron
populations are added at $v_\mathrm{Beam} \approx 5 v_\mathrm{th}$, whose
density is about $10\%$ of the total electron density (see fig.
\ref{fig:setup}). The beam density and velocity is derived from typical Type II
emission region parameters given in \cite{KnockModel}, their pitch angle
distribution is centered around $45^{\circ}$, following \cite{KarlickyVandas}.

In the two-beam case, the simulation will be naturally charge-neutral and have
a zero total current. In the single-beam case however, the background electron
population is suitably boosted in the opposite direction to create an
appropriate return current, and thus also zero out the total current within the
simulation box.

\begin{table}
	\begin{tabular}{lcl}
		\hline
		Physical parameters\\
		\hline
		Background plasma density & $\rho$ & $2.5\cdot 10^7 \,\mathrm{cm^{-3}}$\\
		Temperature               & $T$    & $480280\,\mathrm{K}$\\
		Background magnetic field & $B_x$  & $1\,\mathrm{G}$\\
		Plasma frequency          & $\omega_p$ & $2\cdot10^8\,\mathrm{Rad/s}$\\
		Beam density              & $\rho_B$ & $0.1 * \rho$\\
		\hline
		Numerical parameters\\
		\hline
		Box size in cells & $N_x$ & 8192\\
                      & $N_y$ & 4096\\
		Cell size         & $\Delta x$ & $0.95\,\mathrm{cm}$\\
		Timestep					& $\Delta t$ & $1.2\cdot10^{-11}\,\mathrm{s}$\\
		Particles per Cell& $n$ & 200 Background + 20 Beam\\
		\hline

	\end{tabular}
	\caption{Simlation parameters for the particle in cell simulation}
\end{table}

\section{Results}\label{s:results}

\begin{figure}[htb]
	\centerline{
	  \rotatebox{270}{\includegraphics[width=0.5\textwidth,clip=]{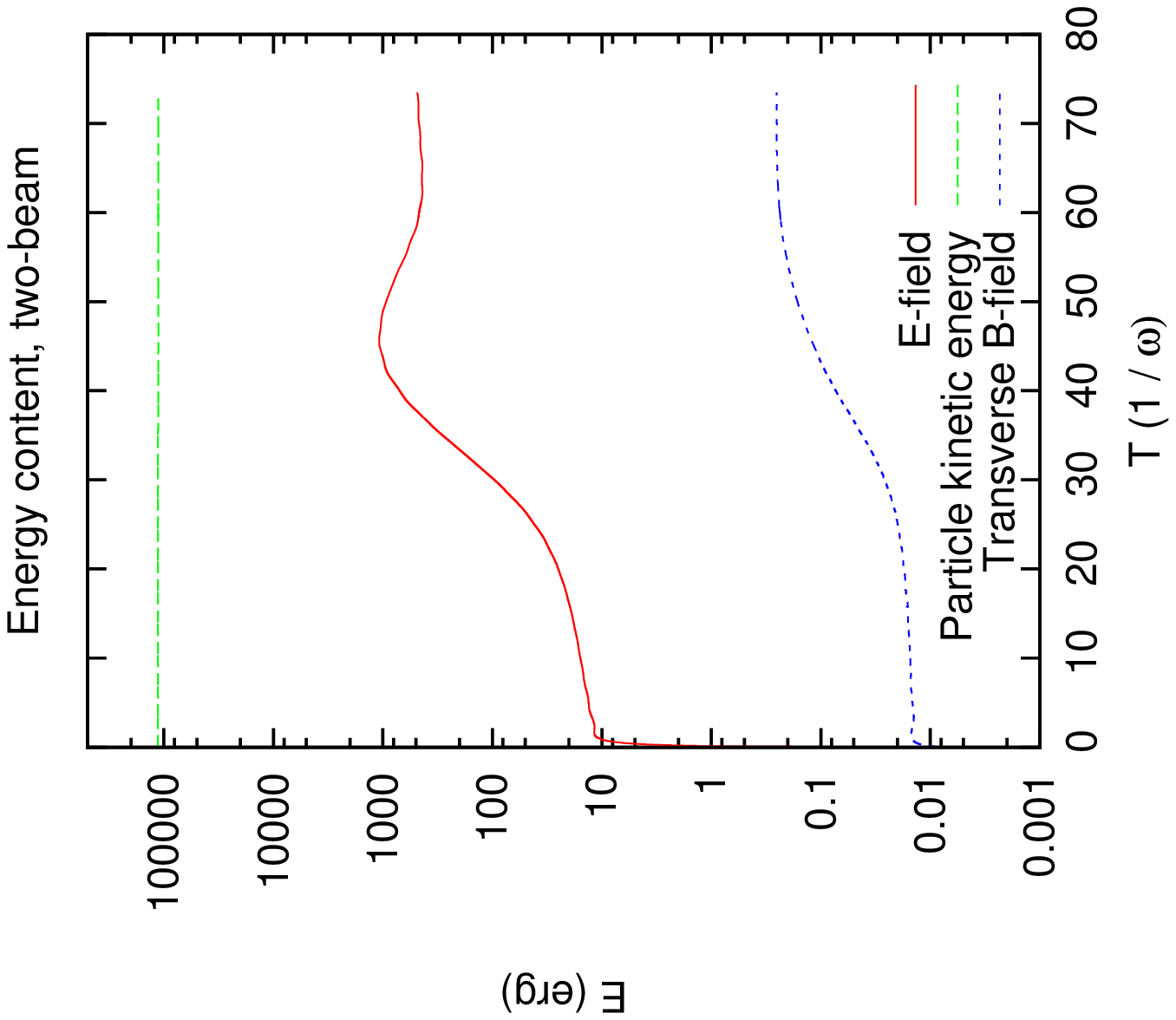}}
	  \rotatebox{270}{\includegraphics[width=0.5\textwidth,clip=]{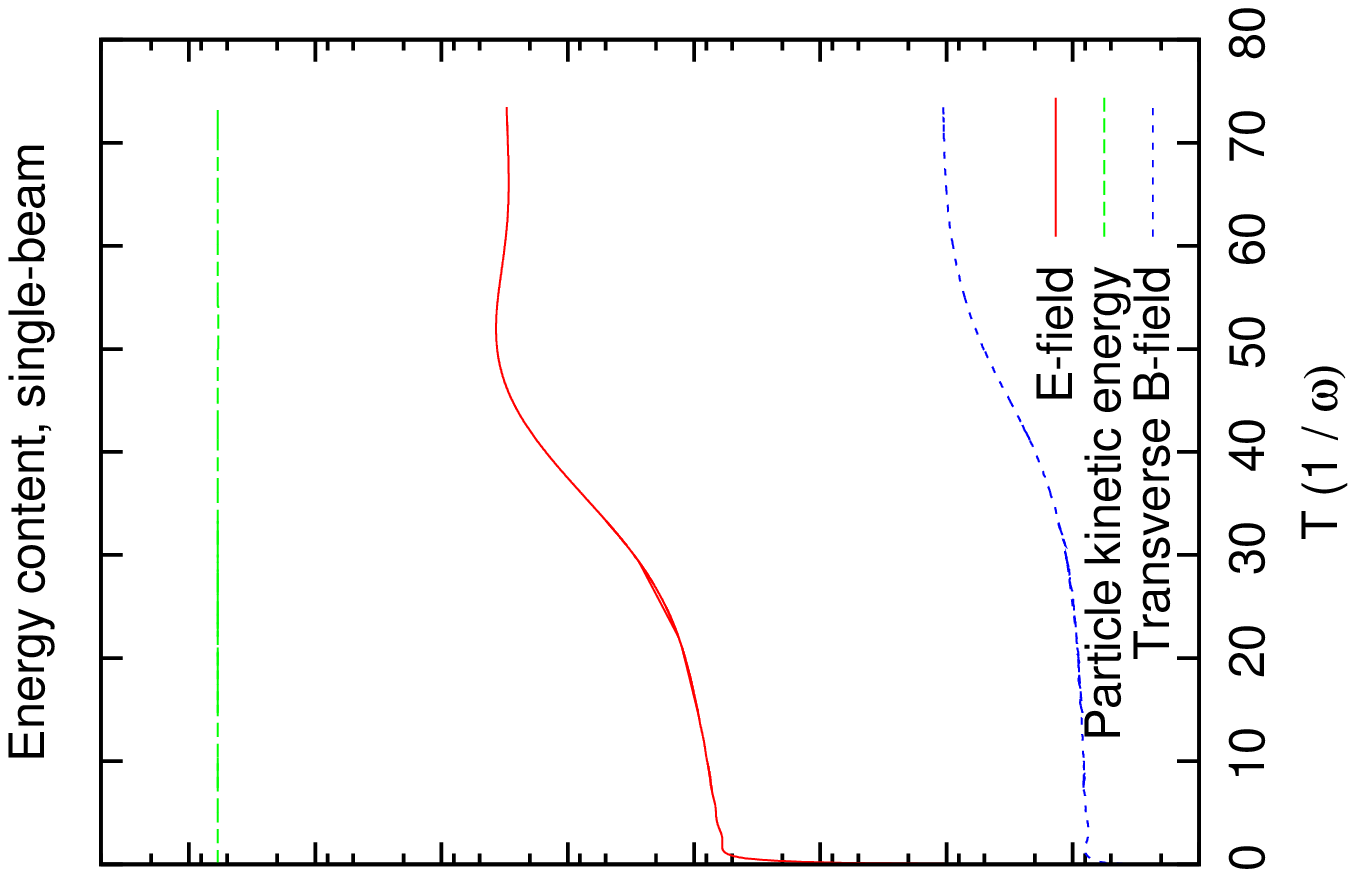}}
	}
	\caption{Energy distribution within two-beam (left) and single-beam (right)
	simulations, over time. After about 20 plasma timescales ($\omega_p ^{-1}$),
	the beam instability causes an increase in electric field energy.
	Subsequently, the transverse magnetic field energy also
	rises. (Background B-Field intensity is not shown, as it is constant
	throughout the simulation)}\label{fig:energyout}
\end{figure}

Since the simulation is started with a quiescent thermal background plasma,
into which the electron beam populations are injected, it takes a while (about $t = 20 \omega_p^{-1}$)  for the
beam-driven instability to create a suitable amount of wave intensity. This can
be easily observed by looking at the energy distribution graphs of single- and
two-beam simulations, which are shown in figure \ref{fig:energyout}.

In this graph, total particle kinetic energy, electric field energy and the
energy content of the transverse magnetic field components is displayed over
simulation time (the background magnetic field component is left out, since it
is constant in good approximation).

The total particle kinetic energy contains a large contribution from the
background plasma's thermal motion, making up about $61\%$ of its value at the
start of the simulation. The remaining $38\%$, or $4.3\cdot10^4\, \mathrm{erg}$
(calculated from the particles' initial velocities) are contribution of the
beam kinetic energy. 

The electric field energy reaches a maximum at $t \approx 50 \omega_p^{-1}$,
with a coinciding rise in energy in the transverse magnetic fields, implying an
energy transfer from electrostatic to electromagnetic modes.

Confirming the results of \cite{KarlickyVandas}, a peak electric field energy
of about $2.8\%$ of the beam kinetic energy is observed in both simulations.
The energy content in the transverse magnetic fields at the end of the
simulation runs lies about 3 orders of magnitude below that of the electric
fields.

\subsection{Single-Beam Scenario}

\begin{figure} 
	\centerline{\rotatebox{270}{\includegraphics[height=0.7\textwidth,clip=]{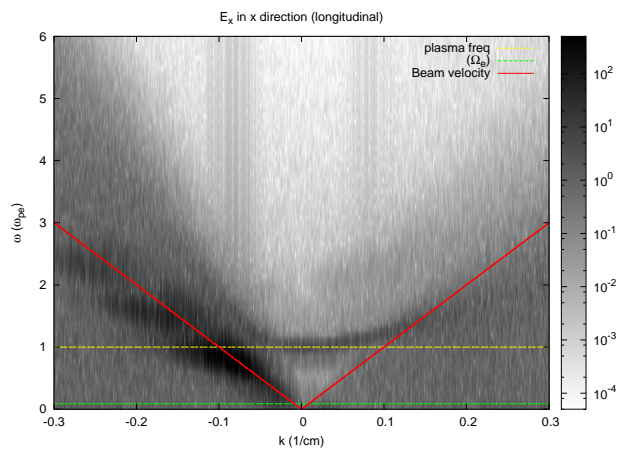}}}
	\caption{Dispersion plot of the longitudinal electric field component along the
	beam direction in a one-beam simulation. A strong excitation of the beam-driven mode, around resonance
	with the plasma frequency is visible in both forward direction (left).}\label{fig:1beam-langmuir}
\end{figure}

In order to obtain quantitative information about the intensities of individual
wave modes within the simulation, electric and magnetic field quantities were
Fourier-transformed in spatial and temporal dimensions. The results are plotted
as $k$ vs.\ $\omega$ intensity plots, in which each wave mode can be identified
via its characteristic dispersion relation.

Figure \ref{fig:1beam-langmuir} shows this dispersion plot for the longitudinal
electric field component ($E_x$) along the beam direction, with $k \parallel
\vec{B}_0$. As expected, electrostatic longitudinal wave modes are discernible in
this plot. In addition to the Langmuir wave, which is forming a parabolic shape
around the plasma frequency at $k = 0$, an additional, strong area of intensity
is visible at negative $k$ values. The dispersion relation of this wave
corresponds to a beam driven mode of the electron beam, which excites
oscillations in the electron beam's density near resonance with the plasma
frequency \citep{WillenGeneralizedLangmuir}.

By taking the sum of the energy content in the forward-propagating half space
and comparing with the with the backward-propagating one, the beam driven
modes' energy can be determined to $E_f = 475\,\mathrm{erg}$, whereas the
backward direction only contains $E_b = 0.3\,\mathrm{erg}$ of energy. Note
however that the energy content determined from this dispersion plot only
contains modes which are exactly axis-parallel (up to numerical precision) and
is therefore grossly underestimating the beam-modes complete energy
contribution (compare figure \ref{fig:energyout}).

\begin{figure}
	\centerline{\rotatebox{270}{\includegraphics[height=0.6\textwidth,clip=]{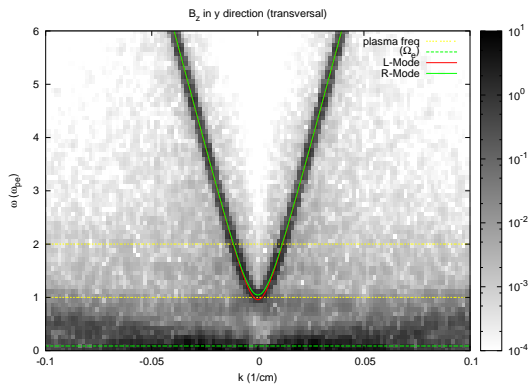}}}
	\caption{Dispersion plot of a transverse magnetic field component in the
	single-beam simulation, with $k$ perpendicular to the beam direction. The
	electromagnetic mode is clearly visible as a parabola, with a cutoff at the
	plasma frequency (splitting into L- and R-Mode is not sufficiently
	numerically resolved). No additional unexpected wave modes are observed.}
	\label{fig:1beam-em}
\end{figure}

In figure \ref{fig:1beam-em}, a similar dispersion plot for a transverse
magnetic field component is shown. Apart from some thermal background
excitation of the electromagnetic mode, no unexpected features are apparent in
this plot. Especially, no indications of three-wave interaction are visible here.

\subsection{Two-Beam Scenario}

\begin{figure}[htb]
	\centerline{\rotatebox{270}{\includegraphics[height=0.6\textwidth,clip=]{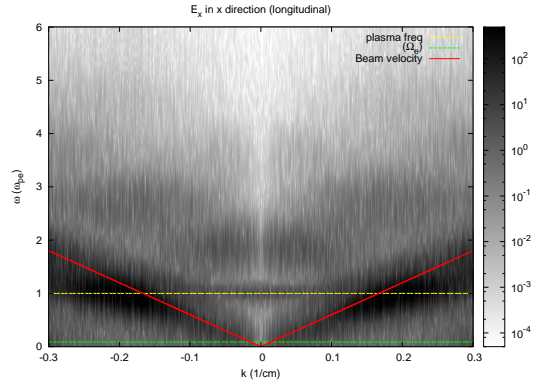}}}
	\caption{Dispersion plot of the longitudinal electric field component along the
	beam direction in a two-beam simulation. A strong excitation of the beam-driven mode, around resonance
	with the plasma frequency is visible in both forward and backward direction.
	Additionally, weaker but significant excitation appears to be present around $2 \omega_p$}\label{fig:2beam-langmuir}
\end{figure}

The longitudinal electric field dispersion plot for the two-beam case (figure \ref{fig:2beam-langmuir}) shows a
symmetric structure of two counterpropagating beam-driven mode peaks, similar to figure \ref{fig:1beam-langmuir}.
Additionally, a somewhat weaker feature is visible around $2 \omega_p$, which
may be indicative of the three-wave coupling processes outlined in section \ref{s:theory}.

\begin{figure} 
	\centerline{\rotatebox{270}{\includegraphics[height=0.6\textwidth,clip=]{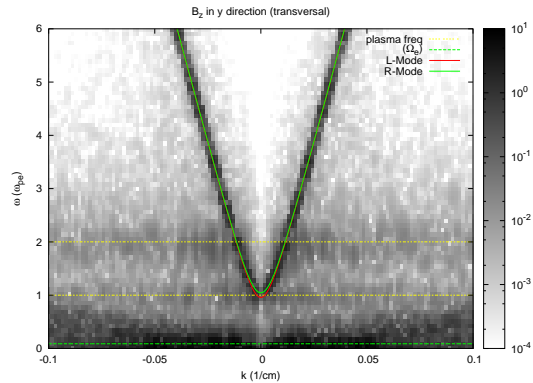}}}
	\caption{Dispersion plot of a transverse magnetic field component in the
	two-beam simulation, with $k$ perpendicular to the beam direction. The
	electromagnetic mode is clearly visible as a parabola, with a cutoff at the
	plasma frequency (splitting into L- and R-Mode is not sufficiently
	numerically resolved).
	In addition, two broad features at the fundamental and harmonic frequency are
	visible, which are in resonance with the em-Mode. (Intensities in arbitrary units)}
	\label{fig:2beam-em}
\end{figure}

The dispersion plots of transverse magnetic field components (figure
\ref{fig:2beam-em}) show the expected dispersion relation of the
electromagnetic mode, as predicted by linear theory. Additional bands of
intensity at the fundamental and harmonic emission frequency are also visible.
Consistent with the predicted processes outlined in section \ref{s:theory},
these bands are likely created by nonlinear interaction of beam-driven waves
with opposite direction of propagation.

For most $k$-values, the nonlinear product wave does not match an eigenmode of
the background plasma and is hence unable to excite a propagating wave, explaining
the low intensity of the feature in the dispersion plot. For $k$-values
matching the electromagnetic mode however, electromagnetic waves can be
resonantly excited, and travel through the plasma.



\section{Conclusion}

We have compared particle in cell simulations of two different microphysical
emission scenarios of type II radio bursts in a CME foreshock settings: one
with a single electron beam accelerated by the shock, and another with two
counterstreaming shock populations of equal strength.

While the single-beam scenario has shown production of electrostatic waves
through beam-driven instabilities which are in line with theoretical
predictions, emission of electromagnetic waves with fundamental and
harmonic frequencies which would be the fingerprint of a type II burst, have
not been observed.

In the two-beam simulation, electrostatic modes are likewise excited, leading
to counterpropagating wave populations. Here, indications of nonlinear wave
interaction are visible in the dispersion diagrams of transverse magnetic field
component, with maxima at the plasma frequency and its first harmonic,
conforming to the theoretically predicted emission process of type II bursts.

\medskip
The results presented here should not be taken as a literary proof that a
multi-beam environment is a necessity for type II emission: The canonical
process to create counterpropagating wave populations for subsequent
interaction is scattering of Langmuir waves on sound waves (eq.  \ref{eq1}). By
using a collisionless particle in cell approach for our simulations, we have
completely neglected the effects of sound waves on the plasma, thus likely
creating somewhat unrealistic wave dynamics. 

Furthermore, due to the large numerical demands of the particle in cell
simulation, the total physical simulation time was limited to a few
microseconds. It is therefore possible that the electrostatic wave population
has not relaxed into an equilibrium condition by the end of the run, such that
scattering effects may have been underrepresented.

Still, the demonstrated strong dependence of harmonic radio emissions on the
existence of counterpropagating electrostatic waves confirms long-standing
theoretical predictions. In plasma environments where this mechanism is of
importance, typical damping time of these waves has to be much larger than
scattering times if a single beam mechanism is to be responsible for the radio
emissions.

\subsection{Outlook and further work}

In this work, only the two extreme cases of a single beam and
counterpropagating of equal strength beams have been investigated. In an
actual type II emission region, beams will likely have different strengths and
pitch angle distributions if they are originating from separate acceleration
regions. The next step will therefore be simulations with varying beam
intensities and angles. 

The parameter space of these simulations further consists of the density / plasma
frequency of the ambient plasma, magnetic field strength and temperature. Given
sufficient computing time, parameters studies in this space may follow suit.

%

%
\begin{acks}
	The authors would like the J\"ulich Supercomputing Centre and the CSC
	Helsinki for their grants of computing time.
	UG and PK acknowledge financial support by the Elite Network of Bavaria.
	FS acknowledges support by the Deutsche Forschungsgemeinschaft, Grand SP1124-1.
	This work has been supported by the European Framework Programme 7 Grant
	Agreement SEPServer - 262773
\end{acks}

%
%
\bibliographystyle{spr-mp-sola}
\bibliography{ursg}  

\begin{thebibliography}{18}
\ifx \bisbn   \undefined \def \bisbn  #1{ISBN #1}\fi
\ifx \binits  \undefined \def \binits#1{#1}\fi
\ifx \bauthor  \undefined \def \bauthor#1{#1}\fi
\ifx \batitle  \undefined \def \batitle#1{#1}\fi
\ifx \bjtitle  \undefined \def \bjtitle#1{\textit{#1}}\fi
\ifx \bvolume  \undefined \def \bvolume#1{\textbf{#1}}\fi
\ifx \byear  \undefined \def \byear#1{#1}\fi
\ifx \bissue  \undefined \def \bissue#1{#1}\fi
\ifx \bfpage  \undefined \def \bfpage#1{#1}\fi
\ifx \blpage  \undefined \def \blpage #1{#1}\fi
\ifx \burl  \undefined \def \burl#1{\textsf{#1}}\fi
\ifx \href  \undefined \def \href#1#2{\textsf{#2}}\fi
\ifx \doiurl  \undefined \def
  \doiurl#1{\href{http://dx.doi.org/#1}{\textsf{#1}}}\fi
\ifx \betal  \undefined \def \betal{\textit{et al.}}\fi
\ifx \binstitute  \undefined \def \binstitute#1{#1}\fi
\ifx \bctitle  \undefined \def \bctitle#1{#1}\fi
\ifx \beditor  \undefined \def \beditor#1{#1}\fi
\ifx \bpublisher  \undefined \def \bpublisher#1{#1}\fi
\ifx \bbtitle  \undefined \def \bbtitle#1{\textit{#1}}\fi
\ifx \bedition  \undefined \def \bedition#1{#1}\fi
\ifx \bseriesno  \undefined \def \bseriesno#1{\textbf{#1}}\fi
\ifx \blocation  \undefined \def \blocation#1{#1}\fi
\ifx \bsertitle  \undefined \def \bsertitle#1{\textit{#1}}\fi
\ifx \bsnm \undefined \def \bsnm#1{#1}\fi
\ifx \bsuffix \undefined \def \bsuffix#1{#1}\fi
\ifx \bparticle \undefined \def \bparticle#1{#1}\fi
\ifx \barticle \undefined \def \barticle#1{}\fi
\ifx \botherref \undefined \def \botherref#1{}\fi
\ifx \url \undefined \def \url#1{\textsf{#1}}\fi
\ifx \bchapter \undefined \def \bchapter#1{}\fi
\ifx \bbook \undefined \def \bbook#1{}\fi
\ifx \bcomment \undefined \def \bcomment#1{#1}\fi
\ifx \oauthor \undefined \def \oauthor#1{#1}\fi
\ifx \citeauthoryear \undefined \def \citeauthoryear#1{#1}\fi
\def \endbibitem {}
\ifx \bconflocation  \undefined \def \bconflocation#1{#1} \fi

\bibitem[\protect\citeauthoryear{Cane, Sheeley, and Howard}{1987}]{Cane1987}
\begin{barticle}
\bauthor{\bsnm{Cane}, \binits{H.V.}},
\bauthor{\bsnm{Sheeley}, \binits{J.}},
\bauthor{\bsnm{Howard}, \binits{R.A.}}:
\byear{1987},
\batitle{Energetic interplanetary shocks, radio emission, and coronal mass
  ejections}.
\bjtitle{J. Geophys. Res.}
\bvolume{92}(\bissue{A9}),
\bfpage{9869}\,--\,\blpage{9874}.
doi:\doiurl{10.1029/JA092iA09p09869}.
\burl{http://dx.doi.org/10.1029/JA092iA09p09869}.
\end{barticle}
\endbibitem

\bibitem[\protect\citeauthoryear{Forbes \textit{et~al.}}{2006}]{WorkingGroupD}
\begin{barticle}
\bauthor{\bsnm{Forbes}, \binits{T.}},
\bauthor{\bsnm{Linker}, \binits{J.}},
\bauthor{\bsnm{Chen}, \binits{J.}},
\bauthor{\bsnm{Cid}, \binits{C.}},
\bauthor{\bsnm{Kóta}, \binits{J.}},
\bauthor{\bsnm{Lee}, \binits{M.}},
\bauthor{\bsnm{Mann}, \binits{G.}},
\bauthor{\bsnm{Mikic}, \binits{Z.}},
\bauthor{\bsnm{Potgieter}, \binits{M.}},
\bauthor{\bsnm{Schmidt}, \binits{J.}},
\bauthor{\bsnm{Siscoe}, \binits{G.}},
\bauthor{\bsnm{Vainio}, \binits{R.}},
\bauthor{\bsnm{Antiochos}, \binits{S.}},
\bauthor{\bsnm{Riley}, \binits{P.}}:
\byear{2006},
\batitle{Cme theory and models}.
\bjtitle{Space Science Reviews}
\bvolume{123}(\bissue{1}),
\bfpage{251}\,--\,\blpage{302}.
\burl{http://dx.doi.org/10.1007/s11214-006-9019-8}.
\end{barticle}
\endbibitem

\bibitem[\protect\citeauthoryear{{H. Aurass}, {B. Vrsnak}, and {G.
  Mann}}{2002}]{TypeIIFlare}
\begin{barticle}
\bauthor{\bsnm{{H. Aurass}}},
\bauthor{\bsnm{{B. Vrsnak}}},
\bauthor{\bsnm{{G. Mann}}}:
\byear{2002},
\batitle{Shock-excited radio burst from reconnection outflow jet?}
\bjtitle{A\&A}
\bvolume{384}(\bissue{1}),
\bfpage{273}\,--\,\blpage{281}.
doi:\doiurl{10.1051/0004-6361:20011735}.
\burl{http://dx.doi.org/10.1051/0004-6361:20011735}.
\end{barticle}
\endbibitem

\bibitem[\protect\citeauthoryear{{Holman} and {Pesses}}{1983}]{Holman1983}
\begin{barticle}
\bauthor{\bsnm{{Holman}}, \binits{G.D.}},
\bauthor{\bsnm{{Pesses}}, \binits{M.E.}}:
\byear{1983},
\batitle{{Solar type II radio emission and the shock drift acceleration of
  electrons}}.
\bjtitle{\apj}
\bvolume{267},
\bfpage{837}\,--\,\blpage{843}.
doi:\doiurl{10.1086/160918}.
\end{barticle}
\endbibitem

\bibitem[\protect\citeauthoryear{{Karlick{\'y}} and
  {Vandas}}{2007}]{KarlickyVandas}
\begin{barticle}
\bauthor{\bsnm{{Karlick{\'y}}}, \binits{M.}},
\bauthor{\bsnm{{Vandas}}, \binits{M.}}:
\byear{2007},
\batitle{{Shock drift electron acceleration and generation of waves}}.
\bjtitle{\planss}
\bvolume{55},
\bfpage{2336}\,--\,\blpage{2339}.
doi:\doiurl{10.1016/j.pss.2007.05.015}.
\end{barticle}
\endbibitem

\bibitem[\protect\citeauthoryear{Kilian, Burkart, and
  Spanier}{2012}]{acronym11}
\begin{bchapter}
\bauthor{\bsnm{Kilian}, \binits{P.}},
\bauthor{\bsnm{Burkart}, \binits{T.}},
\bauthor{\bsnm{Spanier}, \binits{F.}}:
\byear{2012},
\bctitle{The influence of the mass ratio on particle acceleration by the
  filamentation instability}.
In: \beditor{\bsnm{Nagel}, \binits{W.E.}},
\beditor{\bsnm{Kröner}, \binits{D.B.}},
\beditor{\bsnm{Resch}, \binits{M.M.}} (eds.)
\bbtitle{High Performance Computing in Science and Engineering '11},
\bpublisher{Springer},
\blocation{Berlin Heidelberg},
\bfpage{5}\,--\,\blpage{13}.
\bisbn{978-3-642-23869-7}.
doi:\doiurl{10.1007/978-3-642-23869-7}.
\burl{http://dx.doi.org/10.1007/978-3-642-23869-7}.
\end{bchapter}
\endbibitem

\bibitem[\protect\citeauthoryear{{Knock} \textit{et~al.}}{2001}]{KnockModel}
\begin{barticle}
\bauthor{\bsnm{{Knock}}, \binits{S.A.}},
\bauthor{\bsnm{{Cairns}}, \binits{I.H.}},
\bauthor{\bsnm{{Robinson}}, \binits{P.A.}},
\bauthor{\bsnm{{Kuncic}}, \binits{Z.}}:
\byear{2001},
\batitle{{Theory of type II radio emission from the foreshock of an
  interplanetary shock}}.
\bjtitle{Journal of Geophysical Research}
\bvolume{106},
\bfpage{25041}\,--\,\blpage{25052}.
doi:\doiurl{10.1029/2001JA000053}.
\end{barticle}
\endbibitem

\bibitem[\protect\citeauthoryear{Mann}{1995}]{Mann1995}
\begin{bchapter}
\bauthor{\bsnm{Mann}, \binits{G.}}:
\byear{1995},
\bctitle{Theory and observations of coronal shock waves}.
In: \beditor{\bsnm{Benz}, \binits{A.}},
\beditor{\bsnm{Krüger}, \binits{A.}} (eds.)
\bbtitle{Coronal Magnetic Energy Releases}
\bsertitle{Lecture Notes in Physics}
\bseriesno{444},
\bpublisher{Springer Berlin / Heidelberg}, \blocation{???},
\bfpage{183}\,--\,\blpage{200}.
\bcomment{10.1007/3-540-59109-5\_50}.
\bisbn{978-3-540-59109-2}.
\burl{http://dx.doi.org/10.1007/3-540-59109-5\_50}.
\end{bchapter}
\endbibitem

\bibitem[\protect\citeauthoryear{{Melrose}}{1986}]{Melrose}
\begin{bbook}
\bauthor{\bsnm{{Melrose}}, \binits{D.B.}}:
\byear{1986},
\bbtitle{{Instabilities in Space and Laboratory Plasmas}},
\bpublisher{Cambridge University Press},
\blocation{Cambridge, UK.}.
\end{bbook}
\endbibitem

\bibitem[\protect\citeauthoryear{Nel\-son and Mel\-rose}{1985}]{NelsonMelrose}
\begin{bchapter}
\bauthor{\bsnm{{Nelson}}, \binits{G.J.}},
\bauthor{\bsnm{{Melrose}}, \binits{D.B.}}:
\byear{1985},
\bctitle{7}.
\bbtitle{{Type II bursts}},
\bpublisher{Cambridge University Press}, \blocation{???},
\bfpage{333}\,--\,\blpage{359}.
\end{bchapter}
\endbibitem

\bibitem[\protect\citeauthoryear{Pomoell, Vainio, and Kissmann}{2008}]{Pomoell}
\begin{barticle}
\bauthor{\bsnm{Pomoell}, \binits{J.}},
\bauthor{\bsnm{Vainio}, \binits{R.}},
\bauthor{\bsnm{Kissmann}, \binits{R.}}:
\byear{2008},
\batitle{Mhd modeling of coronal large-amplitude waves related to cme
  lift-off}.
\bjtitle{Solar Physics}
\bvolume{253}(\bissue{1}),
\bfpage{249}\,--\,\blpage{261}.
\burl{http://dx.doi.org/10.1007/s11207-008-9186-8}.
\end{barticle}
\endbibitem

\bibitem[\protect\citeauthoryear{Pulupa and Bale}{2008}]{Pulupa2007}
\begin{barticle}
\bauthor{\bsnm{Pulupa}, \binits{M.}},
\bauthor{\bsnm{Bale}, \binits{S.D.}}:
\byear{2008},
\batitle{Structure on interplanetary shock fronts: Type ii radio burst source
  regions}.
\bjtitle{Astrophysical Journal}
\bvolume{676},
\bfpage{1330}\,--\,\blpage{1337}.
\end{barticle}
\endbibitem

\bibitem[\protect\citeauthoryear{Schmidt and
  Gopalswamy}{2008}]{schmidtCMEshocks}
\begin{barticle}
\bauthor{\bsnm{Schmidt}, \binits{J.M.}},
\bauthor{\bsnm{Gopalswamy}, \binits{N.}}:
\byear{2008},
\batitle{Synthetic radio maps of cme-driven shocks below 4 solar radii
  heliocentric distance}.
\bjtitle{Journal of Geophysical Research}
\bvolume{113},
\bfpage{A08104}.
\end{barticle}
\endbibitem

\bibitem[\protect\citeauthoryear{{Spanier} and {Wisniewski}}{2011}]{Wisniewski}
\begin{barticle}
\bauthor{\bsnm{{Spanier}}, \binits{F.}},
\bauthor{\bsnm{{Wisniewski}}, \binits{M.}}:
\byear{2011},
\batitle{{Simulation of Charged Particle Diffusion in MHD plasmas}}.
\bjtitle{Astrophysics and Space Sciences Transactions}
\bvolume{7},
\bfpage{21}\,--\,\blpage{27}.
doi:\doiurl{10.5194/astra-7-21-2011}.
\end{barticle}
\endbibitem

\bibitem[\protect\citeauthoryear{Tsiklauri}{2010}]{Tsiklauri2010}
\begin{botherref}
\oauthor{\bsnm{Tsiklauri}, \binits{D.}}:
2010,
Particle-in-cell, self-consistent electromagnetic wave emission simulations of
  type iii radio bursts.
\textit{Solar Physics}.
\end{botherref}
\endbibitem

\bibitem[\protect\citeauthoryear{Wild and McCready}{1950}]{WildMcCready}
\begin{barticle}
\bauthor{\bsnm{Wild}, \binits{J.}},
\bauthor{\bsnm{McCready}, \binits{L.}}:
\byear{1950},
\batitle{Observations of the spectrum of high-intensity solar radiation at
  metre wavelengths.}
\bjtitle{Australian Journal of Scientific Research}
\bvolume{3}(\bissue{3}),
\bfpage{387}\,--\,\blpage{398}.
\end{barticle}
\endbibitem

\bibitem[\protect\citeauthoryear{{Willes} and
  {Cairns}}{2000}]{WillenGeneralizedLangmuir}
\begin{barticle}
\bauthor{\bsnm{{Willes}}, \binits{A.J.}},
\bauthor{\bsnm{{Cairns}}, \binits{I.H.}}:
\byear{2000},
\batitle{{Generalized Langmuir waves in magnetized kinetic plasmas}}.
\bjtitle{Physics of Plasmas}
\bvolume{7},
\bfpage{3167}\,--\,\blpage{3180}.
doi:\doiurl{10.1063/1.874180}.
\end{barticle}
\endbibitem

\bibitem[\protect\citeauthoryear{{Zlotnik}
  \textit{et~al.}}{1998}]{ThirdHarmonic}
\begin{barticle}
\bauthor{\bsnm{{Zlotnik}}, \binits{E.Y.}},
\bauthor{\bsnm{{Klassen}}, \binits{A.}},
\bauthor{\bsnm{{Klein}}, \binits{K.-L.}},
\bauthor{\bsnm{{Aurass}}, \binits{H.}},
\bauthor{\bsnm{{Mann}}, \binits{G.}}:
\byear{1998},
\batitle{{Third harmonic plasma emission in solar type II radio bursts}}.
\bjtitle{\aap}
\bvolume{331},
\bfpage{1087}\,--\,\blpage{1098}.
\end{barticle}
\endbibitem

\end{thebibliography}

\end{article} 
\end{document}